\documentstyle[11pt,hrd_pasp,twoside,epsfig]{article}
\def\Stromgren{Str{\"o}mgren }

\def\edcomment#1{\iffalse\marginpar{\raggedright\sl#1\/}\else\relax\fi}
\reversemarginpar

\begin{document}
\title{Str{\"o}mgren Color-Temperature Relations for Cool Stars}
\author{James~L.~Clem and Don~A.~VandenBerg}
\affil{Dept.~of~Physics~\&~Astronomy, University~of~Victoria, P.O.~Box~3055, Victoria,~British~Columbia, V8W~3P6, Canada}
\author{Frank~Grundahl}
\affil{Institute~of~Physics~\&~Astronomy, Aarhus~University, 8000~Aarhus~C, Denmark}
\author{Roger~A.~Bell}
\affil{Department~of~Astronomy, University~of~Maryland, College~Park, Maryland, 20742, USA}

\begin{abstract}
A new extensive grid of synthetic Str{\"o}mgren colors and bolometric corrections is presented for F$\,$-$\,$M type dwarf and giant stars having
3000K $\leq$ T$_{\rm{eff}}$ $\leq$ 8000K, 0.0 $\leq$ log $g$ $\leq$ 5.0, and $-$3.0 $\leq$ [Fe/H] $\leq$ 0.5.  These purely theoretical colors are placed 
on the observational system using color calibrations derived from $uvby$ photometry of nearby field stars having $Hipparcos$ parallaxes and 
effective temperatures determined by the infrared flux method.  Modern isochrones transformed to the observed plane using the resultant color-temperature  
relations are able to reproduce the $uvby$ color-magnitude diagrams of several open and globular clusters (notably M$\,$67, M$\,$92, and 
47$\,$Tuc) quite satisfactorily.  

\end{abstract}

\section{Introduction}

The Str{\"o}mgren $uvby$ photometric system offers unique advantages over standard broadband $UBVRI$ photometry since it can provide 
precise estimates of effective temperature, surface gravity, and chemical composition on a star-by-star basis.  However, detailed $uvby$ 
photometric studies for star clusters have suffered in the past from both the lack of precise CCD photometry as well as accurate 
color-temperature (C-T) relations necessary to transform theoretical isochrones and evolutionary tracks to the observational plane.  
Thanks to the recent observing programs of F. Grundahl, a large sample of globular and open clusters have been observed through 
Str{\"o}mgren filters.  This paper presents an attempt to derive a new set of synthetic $uvby$ C-T relations that will hopefully reproduce not only the 
observed colors of field stars, but also yield reliable transformations of isochrones appropriate to old stellar systems.


\section{The Synthetic Str{\"o}mgren Colors: Calculations \& Calibrations}

The new grids of theoretical \Stromgren colors are derived by convolving the $uvby$ filter transmission functions with synthetic 
spectra computed from the latest version of the SSG spectral synthesis code using MARCS stellar atmosphere models as input.  As a 
test of the synthetic colors, we compare a 4 Gyr, solar metallicity isochrone with the $uvby$ CMDs of 
the well studied open cluster M$\,$67.  These initial comparisons show that the isochrone, which has been transformed to the observed 
planes using the raw theoretical colors, fails to provide an adequate fit to the photometric data.  Therefore, a set of first and second order 
color calibrations are derived from a sample of field stars having well-determined estimates of T$_{\rm{eff}}$ from the infrared flux 
method of Blackwell \& Shallis (1977) as well as log$\,g$ and [Fe/H] values (Alonso et al. 1996, 1999 and Houdashelt et al. 2000).  Figure 2 
presents the fits of the 4 Gyr isochrone transformed to various M$\,$67 CMDs using the revised C-T relations produced by applying the 
calibrations illustrated in Figure 1.  Overall the $calibrated$ isochrone given in Figure 2 (solid line) appears to yield a much better agreement 
to the data than its $uncalibrated$ counterpart (dotted line). 

\begin{figure}[hb]
\plotone{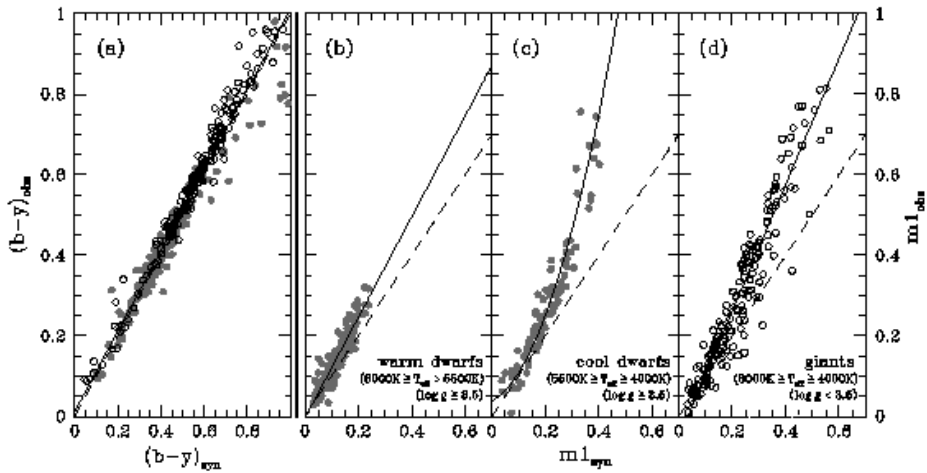}
\caption{ The calibrations required to put the synthetic $(b-y)$ and $m_1$ colors onto the observational systems.  The synthetic 
colors, calculated for each field star with a specific T$_{\rm{eff}}$, log$\,g$, and [Fe/H], are plotted as the abscissae against the 
corresponding observed colors taken from catalog of Hauck \& Mermilliod (1998).  Solid 
lines represent the best fit relations to the data, and dashed 
lines show the one-to-one trends for reference.  The $c_1$ colors (not shown) require only a zero-point correction after the calibrations have been applied 
to the synthetic $(b-y)$ and $m_1$ colors.}  
\end{figure}

\begin{figure}[ht]
\plotone{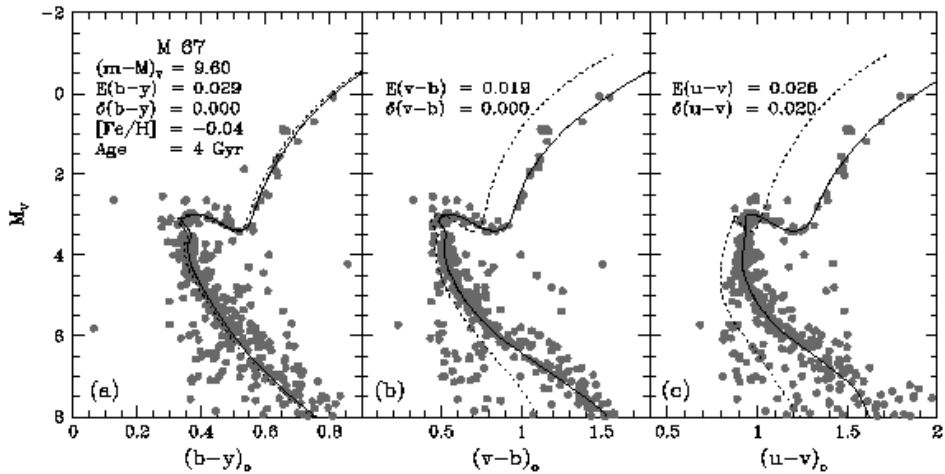}
\caption{ Fits of a 4 Gyr, solar metallicity isochrone to the various $uvby$ CMDs of the open cluster M$\,$67 for the indicated 
parameters.  Solid lines represent isochrones transformed using the newly calibrated C-T relations while the dotted isochrones use the raw, 
uncalibrated colors.  In each panel the color shifts needed to provide the best fits to the data are indicated by the $\delta(b-y)$, 
$\delta(v-b)$, and $\delta(u-v)$ values. }
\end{figure}

\section{Globular Cluster CMDs and the Calibrated \Stromgren Colors}

\Stromgren CMDs of the two prominent, metal-poor globular clusters M$\,$92 and 47$\,$Tuc are used to help constrain the accuracy of the newly 
calibrated $uvby$ C-T relations for sub-solar metallicities.  Figure 3 illustrates the fits of the latest University of Victoria isochrones 
(Bergbusch \& VandenBerg 2001) transformed to the indicated color planes using the calibrated \Stromgren colors derived from the relations 
in Figure 1.  As with M$\,$67, the isochrones yield a superb match to all the pertinent features of the cluster CMDs from the main sequence 
through to the tip of the red giant branch if a short distance scale is assumed.

\section{Conclusions}

In deriving a set of color calibrations necessary to put the new synthetic $uvby$ colors onto the observed systems, we have been able to 
adjust the \Stromgren C-T relations to provide excellent agreement between theoretical model isochrones and the CMDs of such clusters as 
M$\,$67, M$\,$92, and 47$\,$Tuc.  Moreover, isochrone fits to other cluster $uvby$ CMDs not presented here ($e.g.$~NGC$\,$288, M$\,$13, 
NGC$\,$188, NGC$\,$6819, and the Hyades) provide further support for our calibrated C-T relations.  In addition, comparisons of isochrone fits to 
cluster CMDs in the \Stromgren and $UBVRI$ photometric systems yield similar interpretations of the data regardless of which photometric color 
($i.e.$~$b-y$, $B-V$, or $V-I$) is employed.

\newpage
\begin{figure}[h]
\unitlength1.0cm
\begin{center}
\begin{picture}(13.5,6.0)
\put(0,0.50){\epsfig{file=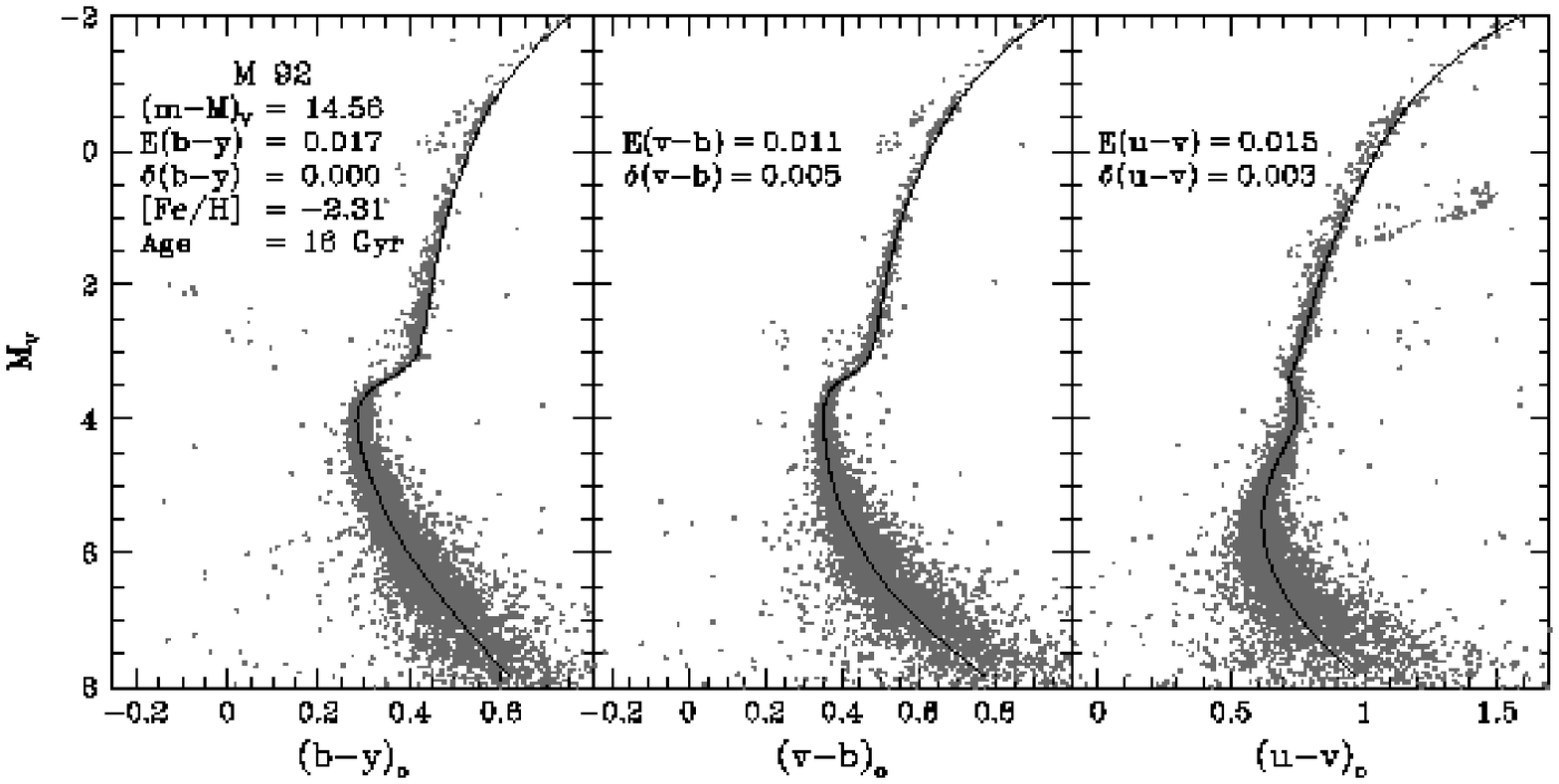,height=6.0cm,width=13.5cm}}
\end{picture}
\end{center}
\end{figure}

\begin{figure}[h]
\unitlength1.0cm
\begin{center}
\begin{picture}(13.5,4.5)
\put(0,0.2){\epsfig{file=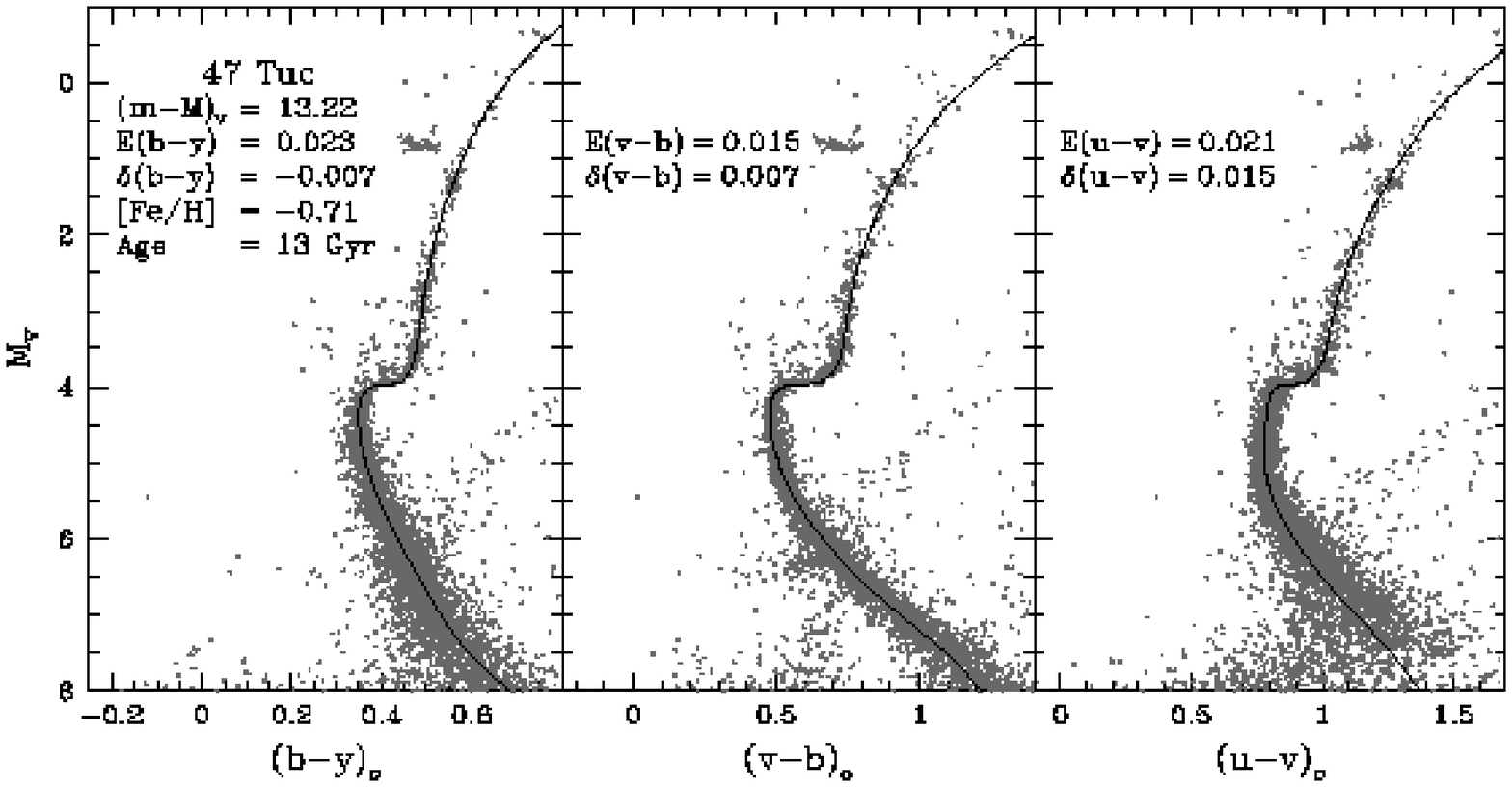,height=6.0cm,width=13.5cm}}
\end{picture}
\end{center}
\caption[]{Isochrone fits to the CMDs of the globular clusters M$\,$92 and 47$\,$Tuc on the assumption of the parameters indicated in the plots.
The isochrones have been transformed to the observed planes using the calibrated \Stromgren C-T relations.}
\end{figure}

\end{document}